\documentclass[12pt]{article}
\usepackage{epsfig}
\usepackage{axodraw
}

\def\beq{\begin{equation}}
\def\eeq{\end{equation}}
\def\bea{\begin{eqnarray}}
\def\eea{\end{eqnarray}}
\def\ba{\begin{array}}
\def\ea{\end{array}}

\def\nn{\nonumber}

\def\Tr{{\rm Tr}}

\def\la{\lambda}
\def\La{\Lambda}
\def\eps{\epsilon}
\def\rd{{\rm d}}
\def\GeV{{\rm GeV}}
\def\eV{{\rm eV}}

\begin{document}

\centerline{\bf\Large Conformal Symmetry and the Standard Model}

\vspace{5mm}
\centerline{\bf Krzysztof A. Meissner${}^{1,2}$ and Hermann Nicolai${}^1$}

\vspace{5mm}
\begin{center}
{\it ${}^1$ Max-Planck-Institut f\"ur Gravitationsphysik
(Albert-Einstein-Institut),\\ M\"uhlenberg 1, D-14476 Potsdam,
Germany\\ ${}^2$ Institute of Theoretical Physics, Warsaw
University,\\ Ho\.za 69, 00-681 Warsaw, Poland}
\end{center}

\begin{abstract}

\footnotesize{We re-examine the question of radiative symmetry
breaking in the standard model in the presence of right-chiral
neutrinos and a minimally enlarged scalar sector. We demonstrate
that, with these extra ingredients, the hypothesis of classically
unbroken conformal symmetry, besides naturally introducing and
stabilizing a hierarchy, is compatible with all available data;
in particular, there exists a set of parameters for which the model
may remain viable even up to the Planck scale. The decay modes of
the extra scalar field provide a unique signature of this model
which can be tested at LHC.}
\end{abstract}
%\maketitle

\noindent
{\bf 1. Introduction.}
A striking property of the standard model (SM) of elementary particles (see
e.g. \cite{N,P} for an introduction and bibliography) is its `near conformal'
invariance. Conformal invariance is only broken by the explicit mass term for
the scalar fields, which induces spontaneous breaking of the
$SU(2)_w \times U(1)_Y$ symmetry and gives mass to the W and Z bosons
\cite{Higgs} as well as to the fermions. This tree level mass term
is also at the root of the so-called hierarchy problem, namely the
need to cancel quadratically divergent terms $\propto \Lambda^2$ to
exceedingly high precision (where $\Lambda$ is the UV cutoff, or the
scale at which the SM is replaced by another theory). The desire to
avoid such unnatural fine tuning, or at least to stabilize such a
seemingly unnatural hierarchy, constitutes the main motivation for
various proposals to extend the SM (see e.g. \cite{Ramond} for a nice
summary), and has in particular led to the development of supersymmetric
extensions of the SM.

Nevertheless, it has been known for a long time that radiative
corrections in an initially conformally invariant scalar field
theory may also induce spontaneous breaking of symmetry, such that
the introduction of explicit mass terms can in principle be
avoided \cite{CW}. However, in spite of its aesthetical appeal,
the concrete implementation of the Coleman Weinberg (CW) mechanism
in the context of the SM so far has not met with much success for
a variety of reasons (but see \cite{Sher,Canadians} for more
recent work). In particular,  we now know that the Higgs mass must
be larger than $115$ GeV, much in excess of the original
prediction ($\sim 10$ GeV) of \cite{CW}, thereby forcing the
scalar self-couplings to be so large that a nearby Landau pole
seems unavoidable and the one-loop approximation may no longer be
valid. As a further constraint, the unexpectedly large Yukawa
couplings of the top quark require the scalar self-coupling to be
sufficiently large in order to prevent the de-stabilization of the
effective potential.

In this letter we re-examine the question of radiative symmetry
breaking for the SM in a slightly more general context than done
before. While the main ingredients that underlie the present work
have been available for a long time, the following three key features
are new\footnote{A similar model but with explicit scalar mass term,
and thus {\em without} radiative symmetry breaking, was recently
proposed and studied in~\cite{Shap}.}.

\begin{itemize}
\item
  We proceed from the hypothesis that classically unbroken
  {\em conformal symmetry} is the basic reason for the existence
  of small mass scales in nature, whose emergence should thus be
  viewed as a manifestation of a {\em conformal anomaly} (other sources
  of violations of conformal invariance in the SM, like gluon and quark
  condensates, are non-perturbative and concern much lower scales than
  we are interested in here).
\item
  {\em Dimensional regularization} is crucial in that it provides a
  self-consistent and economical way to define the renormalized
  effective action to any order in perturbation theory, ensuring that
  conformal invariance is broken `in the least possible way' in the
  quantum theory. However, we stress that its preferred status with
  regard to a Planck scale theory of quantum gravity remains an
  {\em assumption}, see remarks at the end.
\item
  In contrast to previous work on the effective potential, we
  incorporate the right-chiral neutrinos and the associated Yukawa
  (Dirac and Majorana-like) couplings from the outset; this leads us
  to introduce a concomitant scalar field, implementing the standard
  see-saw mechanism \cite{seesaw}. The decay of this scalar fields
  provides a distinctive and unique signature of the present model
  that can be tested at LHC.
\end{itemize}

Our aim is to compute the effective potential for this combined
classically conformally invariant theory, and to derive all known mass
scales from this effective potential. Due to the mixing of the scalar
fields and the presence of logarithmic terms in the effective potential it
now becomes possible to reproduce all the observed features without the
need to introduce unduly large mass hierarchies `by hand'.

Our proposal is `minimalistic' in the sense that we do not invoke
grand unification (GUTs) nor any other `beyond the SM' scenario,
but rely only on those ingredients that are known to be there.
Indeed, the very idea of grand unification, or any other scheme
involving the introduction of a large intermediate scale between
the weak scale and the Planck scale, is evidently at odds with our
basic hypothesis of nearly unbroken conformal invariance --- which is
presumably the reason why the CW potential has not played any role
in GUT scenarios, or softly broken supersymmetric theories (in fact,
as shown in \cite{West} the effective potential vanishes identically
to all orders in an exactly supersymmetric theory). On the other
hand, as we will show here, for a very reasonable range of parameters
these minimal ingredients suffice to reproduce, via the CW mechanism,
all observed features of the SM, including small neutrino masses, in
such a way that Landau poles and instabilities
can be pushed above the Planck scale. Contrary to the usual
reasoning, the smallness of neutrino masses does not necessarily
require a very large `new physics' scale, but can be explained by
the respective neutrino Yukawa couplings if these are taken to be
of the same order as the electron Yukawa coupling ($10^{-5}$). As
our proposal allows for some range of Higgs masses, the (perhaps
sobering) conclusion is that the model proposed here may remain
perfectly viable in all respects well beyond the range of energies
accessible to LHC. In particular, while supersymmetry is expected
to be part of any scheme unifying the basic forces with gravity,
there is no need for {\em low energy} supersymmetry in the present
scheme.\\

\noindent
{\bf 2. Lagrangian and effective potential.}
Omitting kinetic terms the Lagrangian reads (see also \cite{NR},
where (\ref{L}) was considered in the different context of
local Weyl invariance)
\bea
{\cal L}' &=& \Big( \bar{L}^i \Phi Y^E_{ij} E^j +
     \bar{Q}^i \epsilon \Phi^* Y^D_{ij} D^j +
     \bar{Q}^i \epsilon \Phi^* Y^U_{ij} U^j + \nn\\
&&  + \bar{L}^i \epsilon \Phi^* Y^\nu_{ij} \nu^j_R
  + \varphi \nu^{iT}_R {\cal C} Y^M_{ij} \nu^j_R + {\rm h.c.} \Big) \nn\\
&&  - \frac{\la_1}4 (\Phi^\dag \Phi)^2
     - \frac{\la_2}2 \varphi^2 (\Phi^\dag \Phi)
     - \frac{\la_3}4 \varphi^4
\label{L}
\eea
with standard notation: $Q^i$ and $L^i$ are the left-chiral quark
and lepton doublets of $SU(2)_w$, $U^i$ and $D^i$ are the
right-chiral up- and down-like quarks, $E^i$ the right-chiral
`electron-like' leptons, and $\nu_R^i$ the right-chiral neutrinos.
We have suppressed all $SU(2)_w$ and color $SU(3)_c$ indices, but
explicitly indicate family indices $i,j=1,2,3$. The real diagonal
matrices $Y^U_{ij}$, $Y^E_{ij}$, $Y^M_{ij}$ and the complex matrices
$Y^D_{ij}$, $Y^\nu_{ij}$ contain all the relevant Yukawa couplings
and parameterize the most general `family mixing'. Finally,
besides the standard $SU(2)_w$ Higgs doublet $\Phi$, the spectrum
contains an additional real scalar $\varphi$~\footnote{In principle,
 the field $\varphi$ could be taken complex or even to transform in
 a non-trival representation of a family symmetry (in which case $M>1$
 in formula (\ref{Pot1})). The phase of $\varphi$ would then be a
 Goldstone or pseudo-Goldstone boson (sometimes called `Majoron'), which
 couples to observable matter only via the right-chiral neutrinos,
 and might thus be useful for other~purposes.}. Because of the
assumed conformal invariance, no scalar self-couplings other than
those appearing in (\ref{L}) are allowed. The above Lagrangian (together
with the kinetic terms which we have not written) is the most general
compatible with (classical) conformal invariance, and in particular
contains no explicit mass terms (it is also automatically renormalizable).

We next wish to compute the one loop effective (CW) potential.
Using the standard formulas \cite{IZ}, writing $H^2\equiv
\Phi^\dag \Phi$ for the usual Higgs doublet, and defining
\bea
F_\pm (H,\varphi) &:=& \frac{3\la_1 + \la_2}4 H^2
                     + \frac{3\la_3 + \la_2}4 \varphi^2  \pm  \\
   && \!\!\!\!\!\!\!\!\!\!\!\!\!\!\!\!\!\!\!\!\!\!\!\!
        \pm  \sqrt{ \left[\frac{3\la_1 - \la_2}4 H^2 -
                     \frac{3\la_3 - \la_2}4 \varphi^2\right]^2 +
                  \la_2^2 \varphi^2 H^2}   \nn
\label{F}
\eea
the one-loop contributions from the scalar fields to the effective
potential are  (in the $\overline{MS}$-scheme)
\bea
V_{{\rm eff}}^{(1)} (H,\varphi) &=&
  \frac{N-1}{256 \pi^2} \big( \la_1 H^2 + \la_2\varphi^2 \big)^2
  \ln \left(\frac{\la_1 H^2 + \la_2 \varphi^2}{v^2}\right) \nn\\
 &&
\!\!\!\!\!\!\!\!\!\!\!\!\!
+\frac{M-1}{256 \pi^2} \big( \la_2 H^2 + \la_3\varphi^2 \big)^2
  \ln  \left(\frac{\la_2 H^2 + \la_3 \varphi^2}{v^2}\right)  \nn\\
  &&
\!\!\!\!\!\!\!\!\!\!\!\!\!
   + \frac1{64\pi^2} F_+^2 \ln \left(\frac{F_+}{v^2}\right)
  + \frac1{64\pi^2} F_-^2 \ln  \left(\frac{F_-}{v^2}\right)
\label{Pot1}
\eea
where $v$ is some scale (see below).
The formula is valid for $(N+M)$ scalar fields, with $O(N)\times
O(M)$ invariant quartic interactions; in the case at hand, we thus
take $N=4$ (complex doublet) and $M=1$ (real scalar).

With the assumption of classical conformal invariance, it is crucial to
use a regularization for the computation of quantum corrections that
violates this invariance in the least possible way. Unlike other
schemes~\footnote{E.g. with a momentum cutoff $\La$, the conformal
invariance is `more and more badly broken' as $\La\to\infty$.}
dimensional regularization satisfies this requirement
(as it does for ordinary gauge invariance). More explicitly, all
divergent integrals are regulated by the replacement
\beq
\int \frac{d^4 k}{(2\pi)^4} \;\; \longrightarrow \;\;
\frac1{(2\pi v^2)^\epsilon}
\int \frac{d^{4+2\epsilon} k}{(2\pi)^{4+2\epsilon}}
\eeq
where $v$ is some mass scale (which breaks conformal invariance explicitly).
Because $v$ comes with an `evanescent' exponent, {\em the scale
parameter $v$ always appears under a logarithm}. Consequently, the
singular part, and hence the required infinite counterterms are of
the same form as the tree level Lagrangian (\ref{L}), and thus at
any order in perturbation theory, the effective action contains neither
mass terms nor a cosmological constant (which would have to depend
{\em polynomially} on $v$). The one-loop result (\ref{Pot1}) is then
obtained  by analytic continuation of the formula
\beq
\int_0^\infty
\rd\xi\,\xi^{\nu-1}\,\ln(1+b/\xi)=\frac{\pi\,b^\nu}{\nu\,\sin(\pi\nu)}
\eeq
to $\nu=2+\eps$ (the integral converges for $0<{\rm Re} \, \nu <1$).

The computation of the fermionic contribution is more involved due
to family mixing, and cannot be done in closed form without
resorting to some approximations. First of all, inspection of
(\ref{L}) shows that in the one-loop approximation we can separate
the calculation into a part involving only the quark fields, and one
involving only the leptons. The contribution of the quarks is
clearly dominated by the top quark (i.e. the largest Yukawa coupling
$g_t \equiv Y^U_{33}\approx 1.0$) and gives the standard result
\bea
V_{{\rm eff}}^{(2)} (H) = - \frac6{32\pi^2} g_t^4 (H^2)^2 \ln
(H^2/v^2)
\eea
The leptonic contribution, on the other hand, cannot be reduced
so easily as it involves a matrix linking $(L^i,E^i,\nu_R^i)$ and
their charge conjugates. To simplify the calculation, we neglect
all terms involving $Y^E_{ij}$ (whose largest entry comes from the
$\tau$-lepton with $g_\tau$ of order $0.01$). The remaining matrix
only couples the doublets $L^i$ and the right-chiral neutrinos
$\nu_R^i$; before renormalization the relevant expression can be
reduced to the integral
\bea
&& \!\!\!\!\!\!\!\!\!\!\!\!\!\!
-\frac1{16\pi^2(4\pi v^2)^{\eps}\Gamma(2+\eps)} \int d\xi \, \xi^{1+\eps}
\,     \ln\det \bigg[1 + \nn\\
&&  \Big( Y_M Y_M \cdot \varphi^2 + Y_M (Y_\nu\bar{Y}_\nu +
    \bar{Y}_\nu Y_\nu) Y_M^{-1} \cdot H^2 \Big)/ \xi \nn\\
&& \qquad\qquad + \,Y_M \bar{Y}_\nu Y_\nu Y_M^{-1} Y_\nu \bar{Y}_\nu
\cdot (H^2)^2/\xi^2 \bigg]
\eea
where the remaining determinant under the integral is to be taken
w.r.t. a hermitean 3-by-3 matrix in the family indices. Further
evaluation of this expression would thus require the factorization
of a sixth order polynomial in $\xi$ which again is in general not
possible in closed form, especially if there is `maximal mixing'
in the Yukawa matrices (meaning that $Y^\nu_{ij}$ is far away from
a diagonal matrix). For this reason, we resort to yet another
approximation by assuming $Y_\nu\langle H\rangle \!\ll\!
Y_M\langle\varphi\rangle$, in agreement with the observed
smallness of neutrino masses. Then the above expression can be
calculated exactly and the full effective potential becomes, in
this approximation,
\bea
V_{{\rm eff}} (H,\varphi) &=&
\frac{\la_1 H^4}{4}+\frac{\la_2
H^2\varphi^2}{2}+\frac{\la_3\varphi^4}{4}\nn\\
&& \!\!\!\!\!\!\!\!\!\!\!\!\!\!\!\!\!\!\!\!\!\!\!\!  +\frac{3}{256 \pi^2}
\big( \la_1 H^2 + \la_2\varphi^2 \big)^2
  \ln \left[ \frac{\la_1 H^2 + \la_2 \varphi^2}{v^2} \right] \nn\\
  &&  \!\!\!\!\!\!\!\!\!\!\!\!\!\!\!\!\!\!\!\!\!\!\!\!
   + \frac1{64\pi^2} F_+^2 \ln \left[\frac{F_+}{v^2}\right]
  + \frac1{64\pi^2} F_-^2 \ln \left[\frac{F_-}{v^2} \right]
\label{EffPot}\\
&& \!\!\!\!\!\!\!\!\!\!\!\!\!\!\!\!\!\!\!\!\!\!\!\!\!\!\!\!\!\!\!
  - \frac6{32\pi^2} g_t^4 (H^2)^2 \ln\left[ \frac{H^2}{v^2}\right]
- \frac1{32\pi^2} g_M^4 \varphi^4 \ln \left[\frac{\varphi^2}{v^2}\right]
\nn
\eea
where $g_M^4:= \Tr Y_M^4$. We do not include
here the terms from $SU(2)_w\times U(1)_Y$ gauge fields because
the respective gauge couplings are small, nor from $SU(3)_c$ gauge
fields because it is a two-loop effect (although numerically it
can be important and is included in the RG analysis described
below).\\

\noindent
{\bf 3. Minimization of effective potential.}
As we cannot minimize this potential in closed form, we now search
for minima numerically. Since the problem is highly non-linear we
have to use a trial-and-error method in order to arrive at a set of
`reasonable' values satisfying the following requirements: the
standard Higgs mass $m_H$ must be bigger than 115 GeV, and the
effective coupling constants $\la_i^{\rm eff}$ (see (\ref{leff})
below) should be such that there are no Landau poles or instabilities
up to some large scale. The numerical search shows that the `window'
left open by these requirements is not very large, but in particular
allows for the following set of values:
\bea
\la_1=3.4,\; \la_2=2.6,\; \la_3=3.3,\; g_t=1,\; g_M^2 = 0.4.
\label{exemval}
\eea
For these values, the minimum lies at $\langle H\rangle=4.15\cdot 10^{-6}
\, v, \ \langle\varphi\rangle=25.06\cdot 10^{-6}\,v$.
We emphasize that these
numbers are merely chosen to illustrate the possible viability of the
proposed scenario up to very large scales, and by no means constitute
a definitive prediction of our model ({\em idem} for the mass values
(\ref{mHiggs}) below).

Next, we must choose one mass scale which sets the scale for all
other quantities. This we do by imposing $\langle H\rangle=174$
GeV to get in the usual way the observed masses of gauge bosons
and fermions of the Standard Model. Hence,
\beq
\langle H\rangle=174\ \GeV,\;  \langle\varphi\rangle=1050\ \GeV,
\; v=2.41\cdot 10^{5}\langle H\rangle
\eeq
Assuming $|Y_\nu|<10^{-5}$, so the neutrino Yukawa couplings are of
the same order as the electron Yukawa coupling, we can arrive at
very small neutrino masses:
\beq
m_\nu\approx \frac{(Y_\nu\langle
H\rangle)^2}{Y_M\langle\varphi\rangle}< 1\ \eV
\eeq

After symmetry breaking three degrees of freedom of $\Phi$ are
converted into longitudinal components of $W^\pm$ and $Z^0$, so we
are left with two real scalar fields $H$ and $\varphi$, and the
potential $V(\Phi,\varphi)$ should be understood from now as
$V(H,\varphi)$ (with noncanonical normalization). Calculating
second derivatives at the minimum and defining
\bea
H' = H \cos \beta + \varphi \sin \beta \; , \quad
\varphi' = - H \sin \beta + \varphi \cos \beta
\eea
we obtain the mass values:
\beq\label{mHiggs}
m_{H'}=217\ \GeV,\ \ \ \ m_{\varphi'}=439\ \GeV
\eeq
with mixing angle
\beq\label{beta}
 \sin \beta =0.119
\eeq
Note that only the components along $H$ of the mass eigenstates couple
to the usual SM particles. The {\em effective coupling constants} are
calculated as respective fourth-order derivatives of the effective
potential, that is, $\la^{\rm eff}_1 = (1/6)
\partial^4 V_{\rm eff}/\partial H^4$,
{\it etc.}; at the minimum we obtain the values
\beq\label{leff}
\la_1^{{\rm eff}}=1.463\, ,\quad \la_2^{{\rm eff}}=0.348\,, \quad
\la_3^{{\rm eff}}=0.626
\label{laval}
\eeq

\noindent
{\bf 4. Renormalization group analysis.}
The next task is to check for the presence of Landau poles or instabilities
(negative coupling constants) as a function of the scale, and to ascertain
the validity of the one-loop approximation (see e.g.~\cite{Einhorn} for
a discussion of the subtleties involved). Ideally, this would require
calculation of  the full resummed and renormalization group invariant
effective action,
where the Landau pole should manifest itself as a singularity of the
effective momentum dependent terms. However, it appears difficult to
proceed analytically in this way (see \cite{Canadians}) so we content
ourselves here with the conventional procedure, according to which
one should evolve the coupling constants with the renormalization
group equations. Although the initial values are potentially subject
to modification at higher order, we take (\ref{laval}) as the most
natural choice. Defining
$$
y_1=\frac{\la_1^{{\rm eff}}}{4\pi^2} \, , \quad
y_2=\frac{\la_2^{{\rm eff}}}{4\pi^2} \, ,\quad
y_3=\frac{\la_3^{{\rm eff}}}{4\pi^2}\, ,\quad
x=\frac{g_t^2}{4\pi^2} \,,\quad u=\frac{g_M^2}{4\pi^2}
$$
we have the renormalization group equations
\bea
\mu\frac{\rd y_1}{\rd \mu}&=&\frac32 y_1^2+\frac18 y_2^2-6x^2,\nn\\
\mu\frac{\rd y_2}{\rd \mu}&=&\frac38 y_2(2y_1+y_3+\frac43 y_2),\nn\\
\mu\frac{\rd y_3}{\rd \mu}&=&\frac98 y_3^2+\frac12 y_2^2-u^2,
\ \ \ \mu\frac{\rd u}{\rd \mu}=\frac34 u^2\nn\\
\mu\frac{\rd x}{\rd \mu}&=&\frac94 x^2-4xz,\ \ \ \ \mu\frac{\rd
z}{\rd \mu}=-\frac72 z^2.
\label{RG}
\eea
where we added the strong coupling contribution $z=\alpha_s/\pi$.
As dictated by (\ref{EffPot}) we use one-particle-irreducible (and
not the full) $\beta$-functions, since in the effective action the
renormalized external fields are used. With the initial values at
174 GeV given by $g_t=1$, $\alpha_s=0.1$ and (\ref{laval}) one
obtains the evolution curves displayed in Fig.~2, from which it is
evident that there are neither Landau poles nor instabilities below
the Planck scale (the instability occurs at $10^{21}$ GeV). Because
of its non-linearity, the system of coupled evolution equations
(\ref{RG}) is rather delicate and highly sensitive to small changes
in the initial values. Nevertheless, the numerical scan over the
range of parameters satisfying the requirements shows that the
standard Higgs in all cases comes out to be rather light
[$\sim {\cal O} (200 \, \GeV)$], while $m_{\varphi'}$ can vary over a larger
range. However, the determination of the allowed range of values
in the $(m_{H'}, m_{\varphi'})$ parameter plane, which are compatible
with all our requirements and which might lead to more definite predictions,
will be left to future work.\\

\noindent
{\bf 5. Discussion.} Phenomenologically, and for low energies, the
proposed scenario is largely indistinguishable from the SM with
massive neutrinos, but for large energies differs significantly
from SM extensions like the MSSM. Apart from the obvious lack of
superpartners, the Higgs couplings are very different; for instance,
the standard Higgs can now couple to right-chiral neutrinos
via mixing with the new scalar. In fact, the Higgs mixing
provides a rather striking (and unique) signature of the present
model, which would set it apart from other `beyond the SM' scenarios,
and should be testable at LHC~\cite{WB}. Namely, besides decaying
via the usual SM decay modes, the standard Higgs field can now
oscillate into the new scalar $\varphi$, which after reconversion
into $H$ leads to a second resonance at $m_{\varphi'}$ with the same
branching ratios --- thus casting a `shadow' of the standard Higgs
particle, whose size depends on the mixing angle $\beta$, cf. (\ref{beta}).
This process is illustrated in Fig.~1. Unlike for the standard Higgs,
this second resonance can be narrow even for larger values of
$m_{\varphi'}$ if the mixing angle $\beta$ remains sufficiently small.
However, a more detailed discussion of these more phenomenological
aspects is outside the scope of this letter and will be given elsewhere.

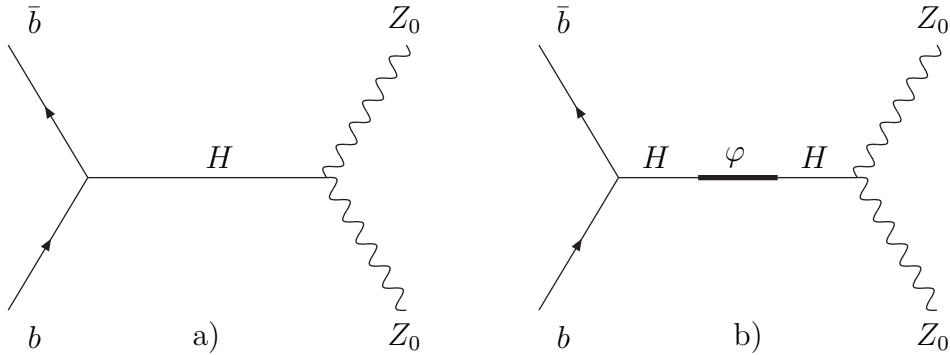
\begin{figure}[htbp]
\begin{center}
%\begin{tabular}{lp{280\unitlength}}
\begin{picture}(370,125)(0,0)
\ArrowLine(10,10)(40,60)
\ArrowLine(40,60)(10,110)
\Line(40,60)(130,60)
\Photon(160,10)(130,60) 3 6
\Photon(130,60)(160,110) 3 6
\Text(90,68)[]{$H$}
\Text(20,0)[]{$b$}
\Text(160,0)[]{$Z_0$}
\Text(20,120)[]{$\bar b$}
\Text(160,120)[]{$Z_0$}
\ArrowLine(210,10)(240,60)
\ArrowLine(240,60)(210,110)
\Line(240,60)(270,60)
\SetWidth{2}
\Line(270,60)(300,60)
\SetWidth{0.5}
\Line(300,60)(330,60)
\Photon(360,10)(330,60) 3 6
\Photon(330,60)(360,110) 3 6
\Text(255,68)[]{$H$}
\Text(285,68)[]{$\varphi$}
\Text(315,68)[]{$H$}
\Text(220,0)[]{$b$}
\Text(360,0)[]{$Z_0$}
\Text(220,120)[]{$\bar b$}
\Text(360,120)[]{$Z_0$}
\Text(85,0)[]{a)}
\Text(290,0)[]{b)}
\end{picture}
\end{center}
\caption{Production of $Z_0$ pairs in a) Higgs and b) new scalar
channels (resonant production is possible if $m_H>2m_Z$ or
$m_\varphi>2m_Z$)}
\label{fig:scalar}
%\end{tabular}
\end{figure}

It is worthwhile to note that the usual hierarchy problem is
addressed here in a way which is very different from the solution
proposed in the context of the MSSM (see e.g. \cite{Weinberg}).
The latter relies mainly on the fact that supersymmetry forces the
Higgs self-coupling to be a function of the gauge couplings, which
themselves are kept under control by gauge invariance, so the Landau
pole effectively gets shifted beyond the Planck scale. The hierarchy
itself is explained in the MSSM by certain soft supersymmetry breaking
terms extremely finely tuned at the GUT scale that run slowly in such
a way that $m_H^2$ eventually becomes negative around 1~TeV.
While the absence of quadratic and quartic divergencies in supersymmetric
theories is due to boson fermion cancellation, there are, by contrast,
neither mass terms nor a cosmological constant in our proposal by
virtue of the assumed conformal invariance. The theory contains
only dimensionless parameters to start with, and the dimensional
regularization ensures that the conformal symmetry is broken in the
radiative corrections not by powers of the cutoff, but only
by the unavoidable choice of a scale $v$ under the logarithms.
In this sense the hierarchy of scales emerges more `naturally' than
it would with explicit mass terms. Although the Landau pole or
instability problems are in principle there, they can be avoided
without excessive fine-tuning as we showed.

The key question is therefore how a classically conformally invariant
action at low energies can emerge from gravity, which is {\em not}
conformally invariant due to the presence of a dimensionful parameter,
the Planck mass $M_{Pl}$ (with gravity and explicit scalar mass terms,
classical dilatational invariance --  achieved by means of a dilaton --
is likewise broken by quantum effects, see \cite{BB}). More precisely,
can the privileged status of dimensional regularization be explained
by the fact that a finite theory of quantum gravity must act as a
universal regulator for matter interactions? If so, quantum gravity
effects may {\em dynamically} suppress explicit breaking of conformal
invariance by powerlike counterterms in this way, allowing only for
logarithmic terms or non-local terms with {\em inverse} powers of
$M_{Pl}$ in the effective action~\footnote{The suppression
of powerlike terms is also suggested by the fact that, whatever the
correct theory of quantum gravity will turn out to be, it must be
such that gravity smoothly decouples in the limit $M_{Pl} \to\infty$ so
as to leave a flat space quantum field theory.}.
A possible analog for such a phenomenon is noncritical
string (Liouville) theory \cite{KPZ}, a theory of matter-coupled quantum
gravity in two space-time dimensions, which does not possess
classical conformal invariance, but where conformal invariance is
restored at the quantum level via the quantum mechanical
decoupling of an infinite tower of null states (as explained e.g.
in \cite{zjkam}).

\vspace{4.0mm}
\noindent
{\bf Acknowledgments:} We are most grateful to J.D.~Bjorken, W.~Buchm\"uller,
T.~Damour, M.~Einhorn, M.~Shaposhnikov and M.~Shifman for critical comments
and discussions. K.A.~Meissner thanks the Albert-Einstein-Institut
for hospitality and financial support.

\newpage

\vspace{-2cm}

\begin{figure}[h]

$
\begin{array}{cc}
        \epsfxsize=7cm
        \epsffile{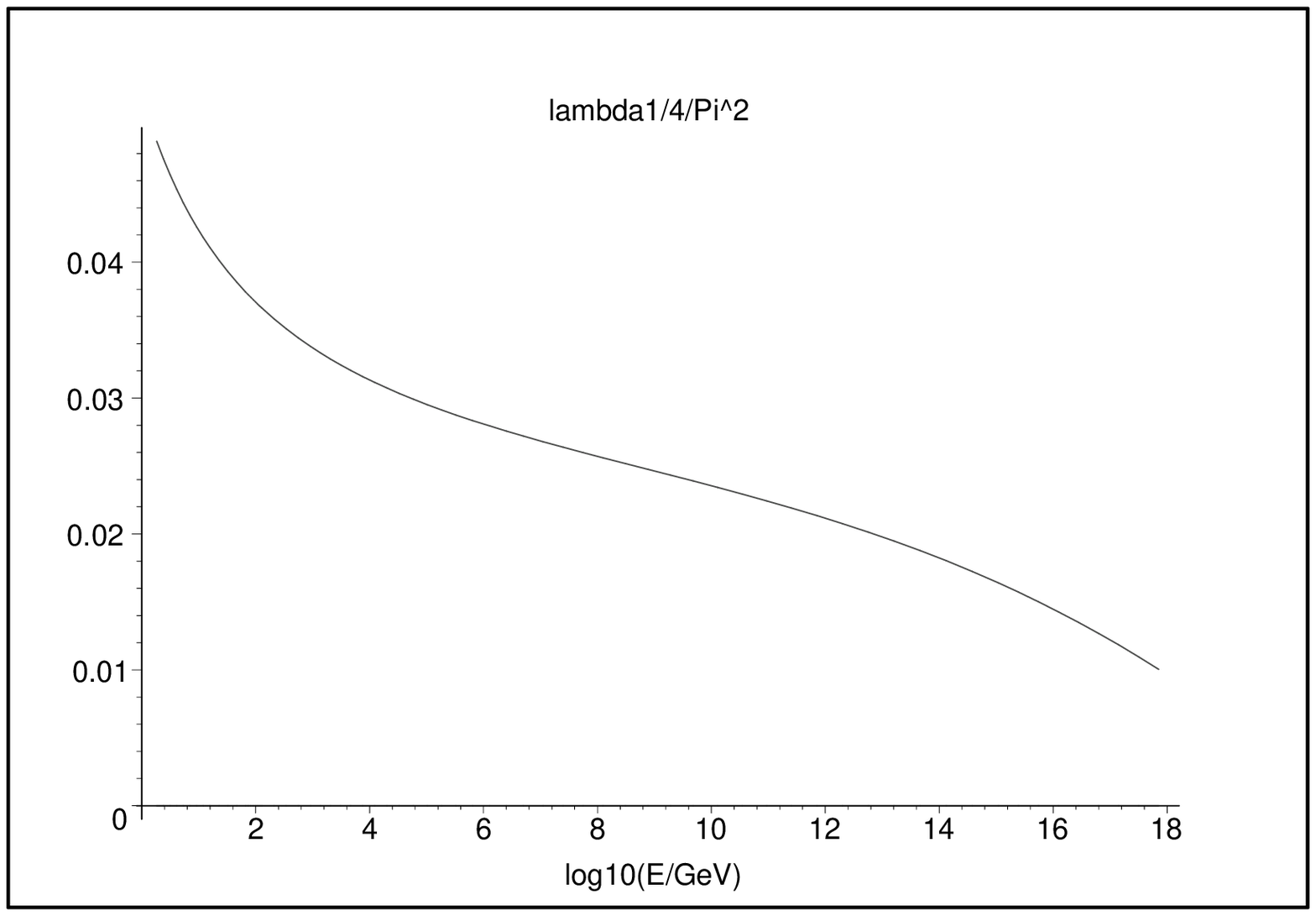} &
         \epsfxsize=7cm
        \epsffile{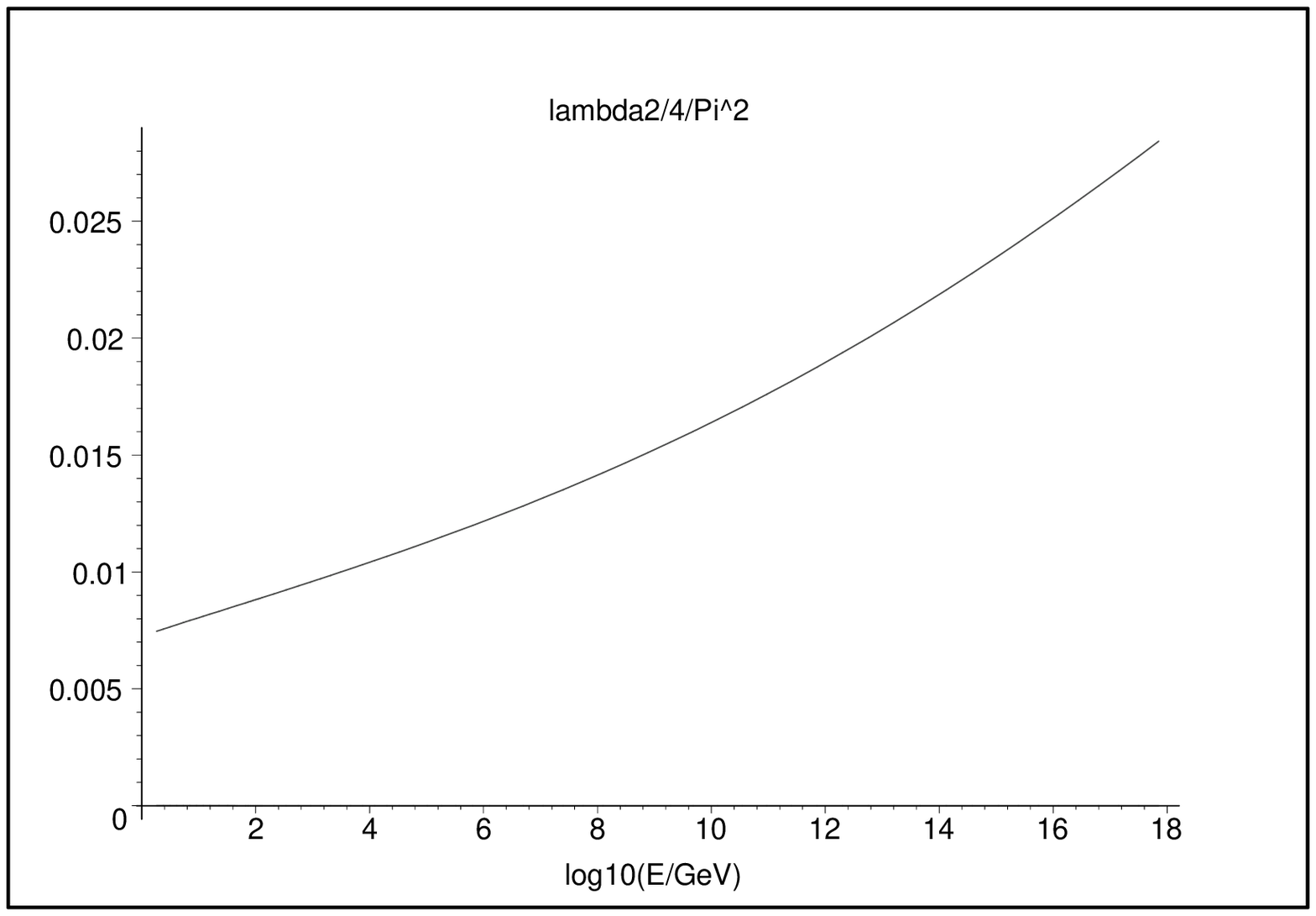} \\[5mm]
         \epsfxsize=7cm
        \epsffile{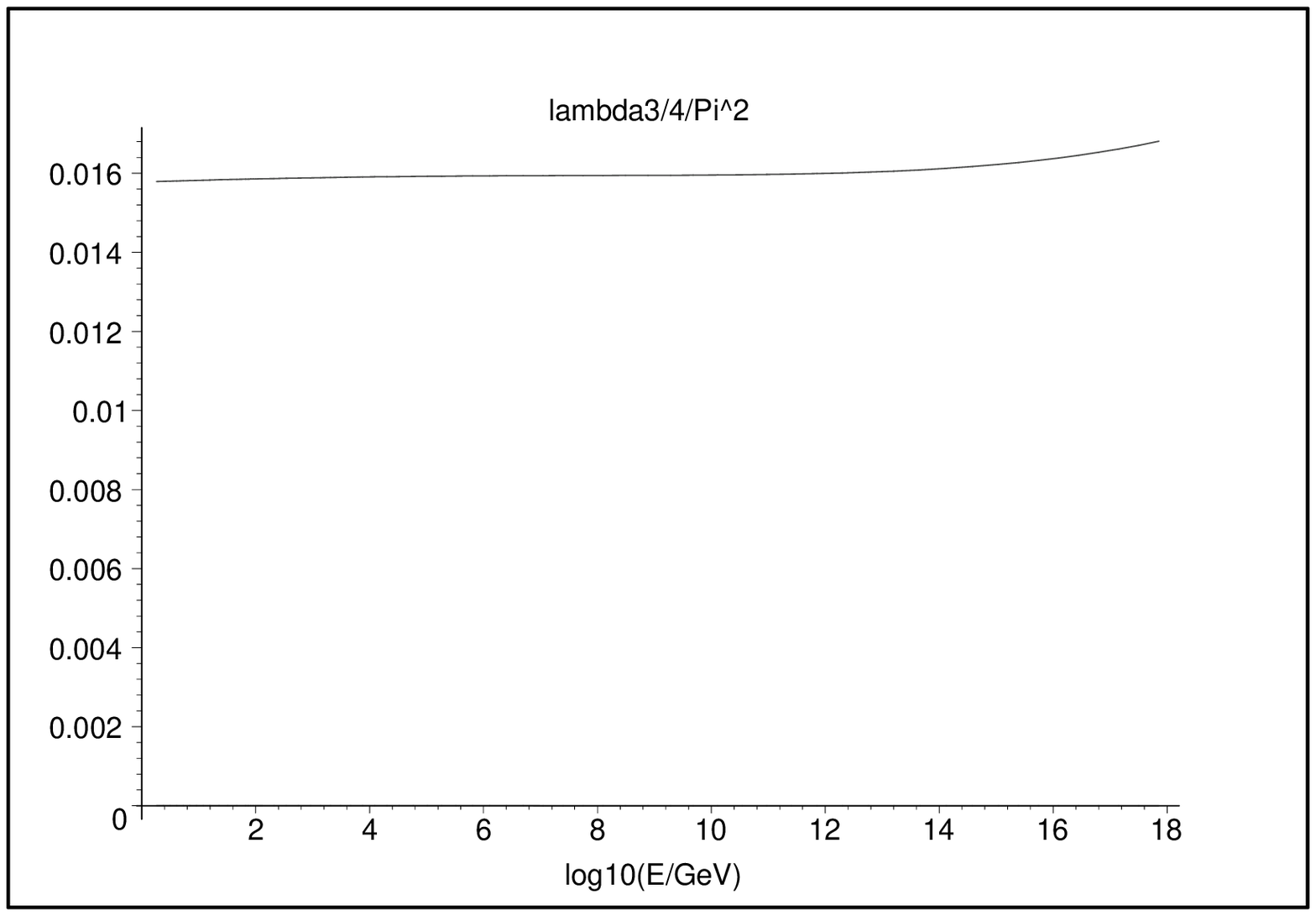}&
         \epsfxsize=7cm
        \epsffile{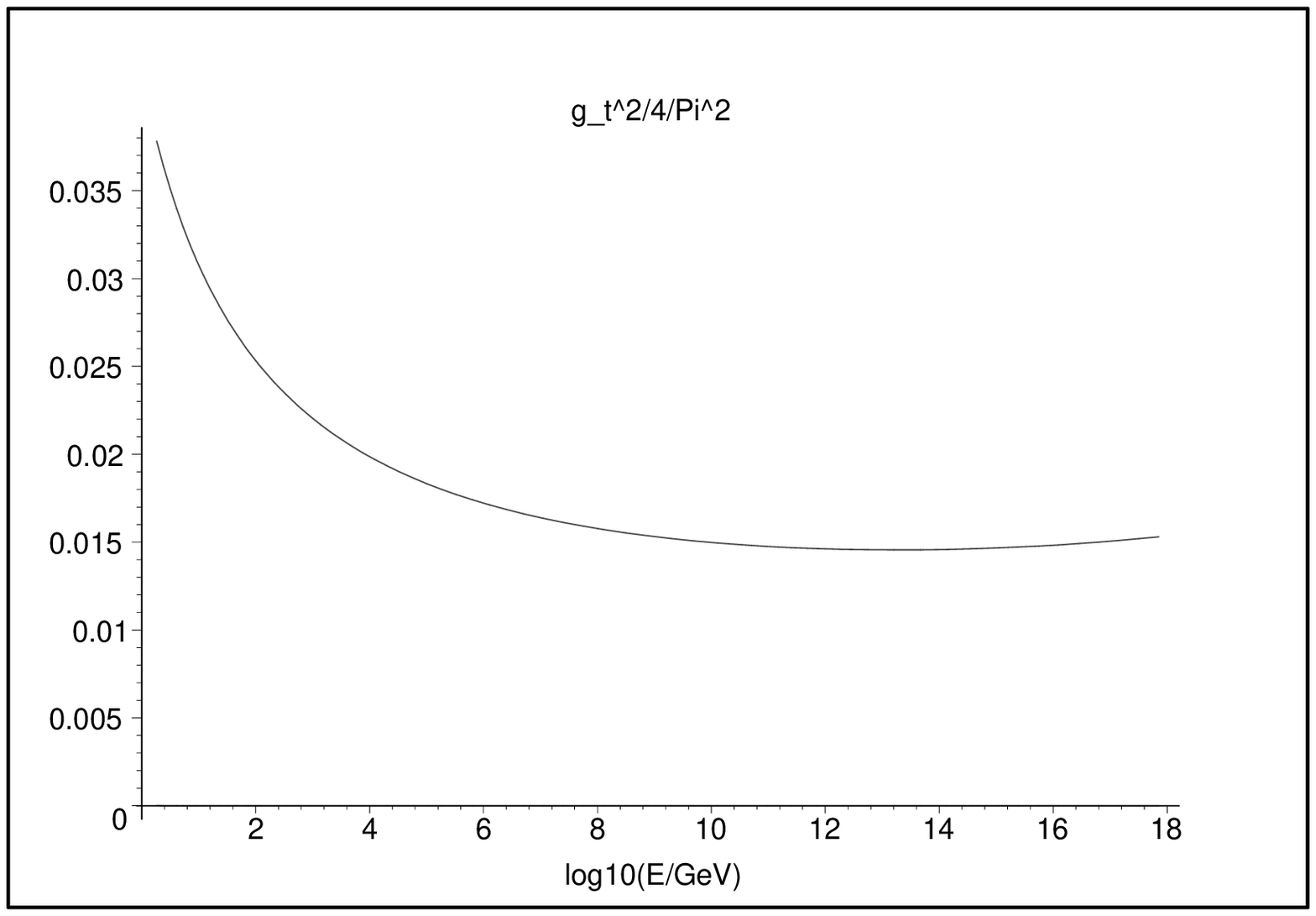} \\[5mm]
         \epsfxsize=7cm
        \epsffile{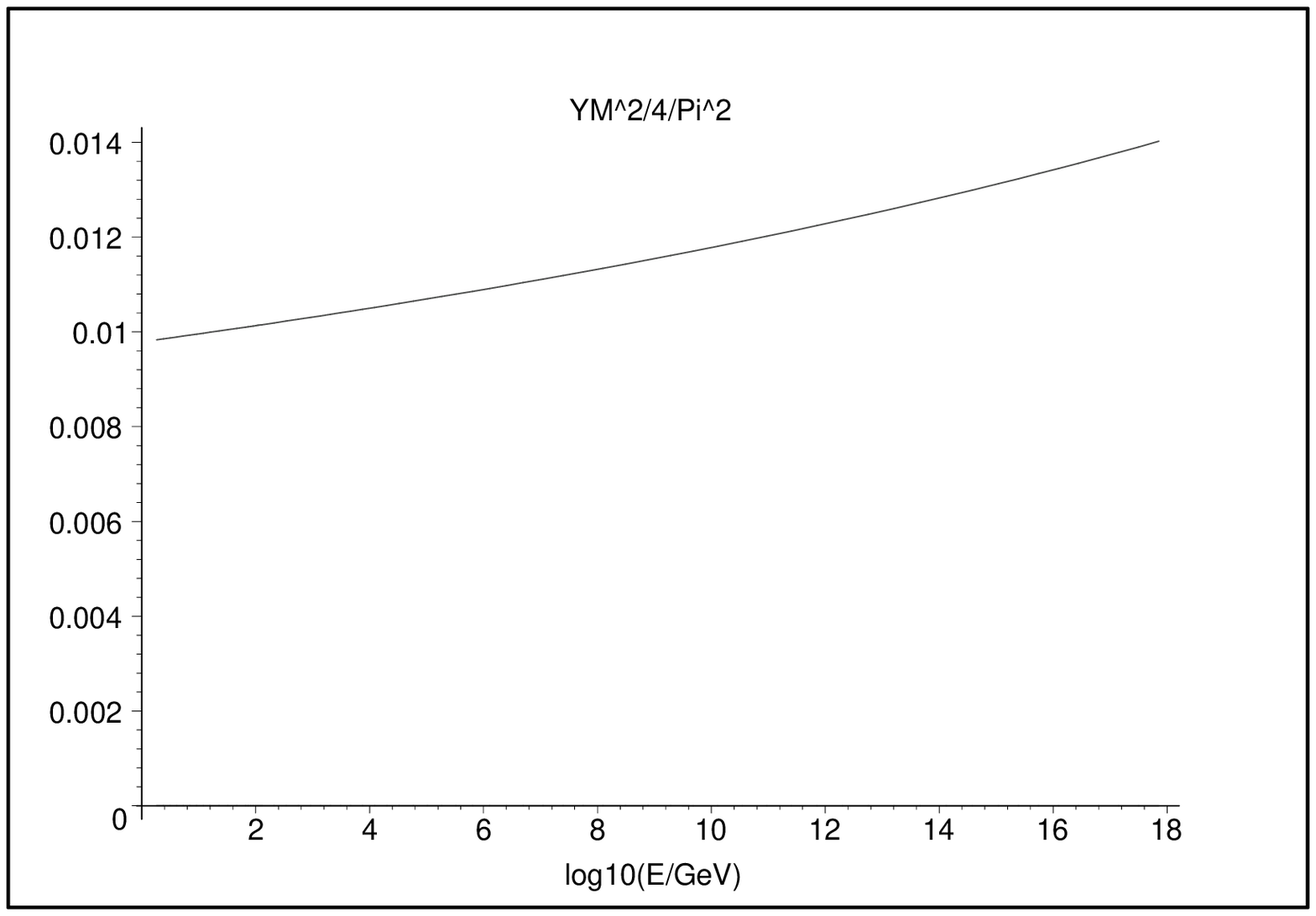} &
         \epsfxsize=7cm
        \epsffile{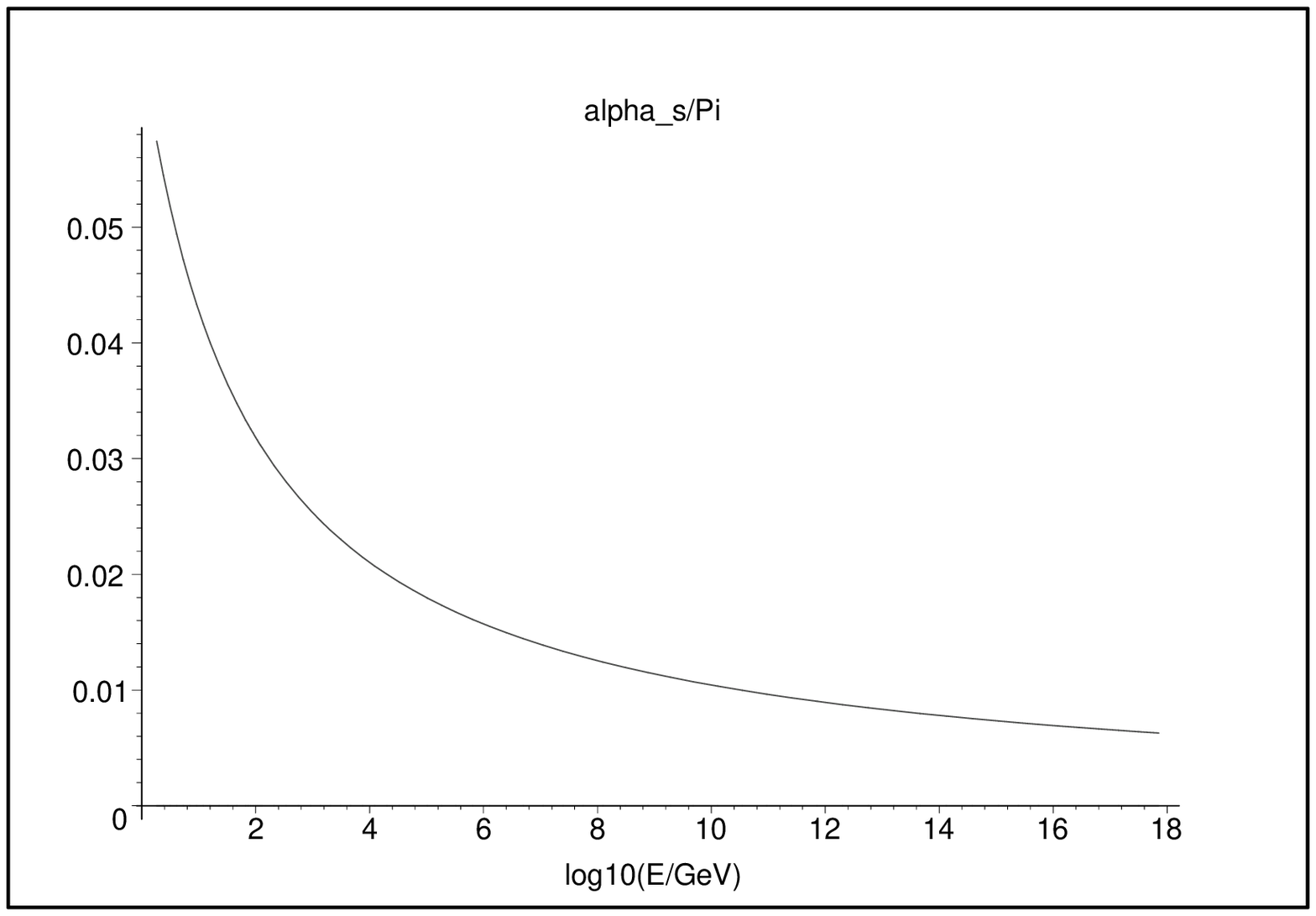}
\end{array}$
\caption{Evolution of coupling constants (from left to right, and top to
bottom): $y_1(\mu),\ y_2(\mu),\ y_3(\mu),\ x(\mu),\ u(\mu)$ and $z(\mu)$}

\end{figure}


\begin{thebibliography}{Ref}
\bibitem{WB} We are indebted to W.~Buchm\"uller for a discussion on
 this point.
\end{thebibliography}
\end{document}